\documentclass[aps,prl,twocolumn,superscriptaddress,showpacs,floatfix]{revtex4}
\usepackage{graphicx}
\usepackage{hyperref}
\usepackage{amsmath}
\usepackage{amssymb}
\usepackage{color}

\newcommand{\ket}[1]{|{#1}\rangle}
\newcommand{\bra}[1]{\langle #1 | }

\newcommand{\Tr}{\mathrm{Tr}}
\newcommand{\mean}[1]{\langle #1 \rangle}

\newcommand{\shp}{\sigma_\mathrm{H}^\dagger}
\newcommand{\sh}{\sigma_\mathrm{H}}
\newcommand{\svp}{\sigma_\mathrm{V}^\dagger}
\newcommand{\sv}{\sigma_\mathrm{V}}

\begin{document}
\flushbottom
\title{Enhanced two-photon emission from a dressed biexciton}

\author{Carlos S\'anchez Mu\~noz}
\affiliation{Departamento de F\'isica Te\'orica de la Materia Condensada
  and Condensed Matter Physics Center (IFIMAC), Universidad Aut\'onoma
  de Madrid, 28049, Spain.}
\author{Fabrice P. Laussy}
\affiliation{Departamento de F\'isica Te\'orica de la Materia Condensada
  and Condensed Matter Physics Center (IFIMAC), Universidad Aut\'onoma
  de Madrid, 28049, Spain.}
\affiliation{Russian Quantum Center, Novaya 100, 143025 Skolkovo, Moscow Region, Russia}
\author{Carlos Tejedor}
\affiliation{Departamento de F\'isica Te\'orica de la Materia Condensada
  and Condensed Matter Physics Center (IFIMAC), Universidad Aut\'onoma
  de Madrid, 28049, Spain.}
\author{Elena del Valle}
\affiliation{Departamento de F\'isica Te\'orica de la Materia Condensada
  and Condensed Matter Physics Center (IFIMAC), Universidad Aut\'onoma
  de Madrid, 28049, Spain.}
\email{elena.delvalle.reboul@gmail.com}
%\affiliation{}

\begin{abstract}
  Radiative two-photon cascades from biexcitons in semiconductor
  quantum dots under resonant two-photon excitation are promising
  candidates for the generation of photon pairs.  In this work, we
  propose a scheme to obtain two-photon emission that allows to
  operate under very intense driving fields. This approach relies on
  the Purcell enhancement of two-photon virtual transitions between
  states of the biexciton dressed by the laser. The richness provided
  by the biexcitonic level structure allows to reach a variety of
  regimes, from antibunched and bunched photon pairs with polarization
  orthogonal to the driving field, to polarization entangled
  two-photon emission. This evidences that the general paradigm of
  two-photon emission from a ladder of dressed states can find
  interesting, particular implementations in a variety of systems.
 \end{abstract}
\pacs{42.50.Pq, 03.67.Bg, 78.67.Hc, 42.50.Dv} \date{\today} \maketitle
\section{Introduction}

The generation of non-­classical states of light is a major goal in
the implementation of photonic quantum
technologies~\cite{obrien09a,knill01a}. A case of particular interest
is the generation of photon pairs, since they present a wide range of
applications in quantum information and quantum
communications~\cite{pan12a}. Photon pairs are an important resource
to generate heralded single photons~\cite{hong86a} and are also used
as a key element for quantum key
distribution~\cite{jennewein00a,naik00a}, quantum
teleportation~\cite{bouwmeester97a,marcikic03a} or to implement
entanglement swapping and quantum
repeaters~\cite{pan98a,simon07a,troiani14a}. Numerous other examples,
like quantum lithography~\cite{giovannetti04a}, the absorption rate
increase from organic molecules in two-photon
microscopy~\cite{geabanacloche89a,upton13a}, quantum walks of
correlated photons~\cite{peruzzo10a} or the quantum computation of
molecular properties~\cite{lanyon10a}, illustrate the rich variety of
applications that these non-classical states of light can find.

Despite having such an impressive number of applications, the number
of ways in which these states can be generated is limited: most
sources of photon pairs employed to date are based on parametric
down-conversion~\cite{burnham70a,kwiat95a,bouwmeester97a,lanyon10a,pan12a}.
This mechanism can be implemented in several platforms, and thankfully
for prospective technologies, semiconductors are demonstrating
excellent performances~\cite{lanco06a,horn12a,boitier14a}.  These
sources suffer however from the major drawback that the number of
photon pairs generated in each process shows Poissonian statistics,
with a non-zero probability of having zero or more than one
pair~\cite{scarani05a}.
Promising candidates to overcome this difficulty are quantum emitters
that naturally emit entangled photon pairs in a radiative
cascade~\cite{aspect81a}. In the semiconductor case, the biexciton
$\ket{\mathrm{B}}$ in a quantum dot offers such an implementation,
and emission of entangled photon pairs from the biexcitonic cascade
has been demonstrated in recent
years~\cite{stevenson06a,akopian06a,hafenbrak07a,dousse10a,muller14a}.
As an alternative to off-resonant excitation, it is possible to
initialize the biexcitonic state by coherent two-photon excitation
(TPE)~\cite{flissikowski04a,stufler06a,jayakumar13a,muller14a,arxiv_huber15a}, which
increases the coherence and indistinguishability of the emitted
photons as compared to non-resonant pumping. The generation efficiency
and the indistinguishability of the photons can also be improved by
bringing a cavity in resonance with the biexcitonic
transition~\cite{dousse10a} to enhance the emission thanks to the
Purcell effect~\cite{birowosuto12a}. A particularly interesting
possibility is to place the cavity in resonance with half the energy
of the biexciton to enhance the rate of spontaneous two-photon
emission, such that two photons are emitted simultaneously into the
cavity mode~\cite{delvalle10a,ota11a,schumacher12a}. The joint
implementation of coherent excitation and Purcell enhancement via
cavity modes has already been discussed in the literature and
shown to be promising~\cite{ota11a,muller14a}.

Under coherent excitation, the intensity of the pumping sets a limit
to the repetition rate of two-photon generation, since strong driving
fields dress the excitons and spoil the biexcitonic
structure~\cite{cohentannoudji77a,muller08a}. On the other hand, a
recent proposal~\cite{sanchezmunoz14a} took advantage of such a
dressing and demonstrated that a continuous source of $N$-­photon
states---with photon pairs as the simplest realization---can be
achieved in precisely this regime of strong admixing of the exciting
laser with the emitter. This relies on a family of virtual two-photon
transitions, so-called \emph{leapfrog} processes~\cite{delvalle13a}, in the
strong driving of resonance fluorescence. Since virtual two-photon
states are emitted away from the fluorescence peaks, they have a very
small probability to occur on their own. These elusive photons are
however precious~\cite{gonzaleztudela15a} since they feature giant
quantum correlations and violate classical
inequalities~\cite{sanchezmunoz14b}. Despite their scarcity, their
existence has recently been demonstrated experimentally by
measurements of frequency-resolved photon
correlations~\cite{peiris15a}.  It is therefore timely to capitalize on
these photons to devise bright, continuous sources of photon pairs by
harvesting them with a cavity mode, with a Purcell-effect applied
similarly to previous enhancements of quantum
correlations~\cite{quang93a,kim14a} from real photons emitted at the
sidebands~\cite{aspect80a,schrama92a,ulhaq12a}.

In this work, we bring together the three main ideas exposed above: i)
TPE from the biexciton, ii) cavity Purcell-enhancement of virtual
processes and iii) multiphoton emission from a dressed system. This
realizes a versatile two-photon source operating in the continuous
regime with a high repetition rate. In comparison with the case of a
single dressed two-level system, the biexciton introduces an extra
degree of freedom, the polarization, that provides a richer set of
physical regimes. In particular, we demonstrate the emission of
degenerate photon pairs with polarization orthogonal to the
laser---therefore suppressing the laser background and undesired
excitation of the cavity---and different two-photon counting
statistics, as well as emission of polarization-entangled
photons.
All these different regimes can be accessed optically with
the same sample just by changing the intensity and polarization of the
excitation. This unprecedented versatility will push forward the
generation and use of photon pairs in the laboratory. Even more
importantly, it evidences that the fundamental concepts are
susceptible to be applied in different platforms, such as
superconducting circuits~\cite{baur09a}, and that new regimes of
non-classical light emission are within reach with variations of the
design.

Our analysis starts with a general introduction of the model and
follows with a detailed description of the features of the dressed
biexciton alone, to finally move to the complete picture with the
inclusion of a cavity that probes and enhances the single and
two-photon transitions present in the dressed system.

\section{Model and dressed state picture}

\begin{figure}[t]
\begin{center}
\includegraphics[width=0.8\linewidth]{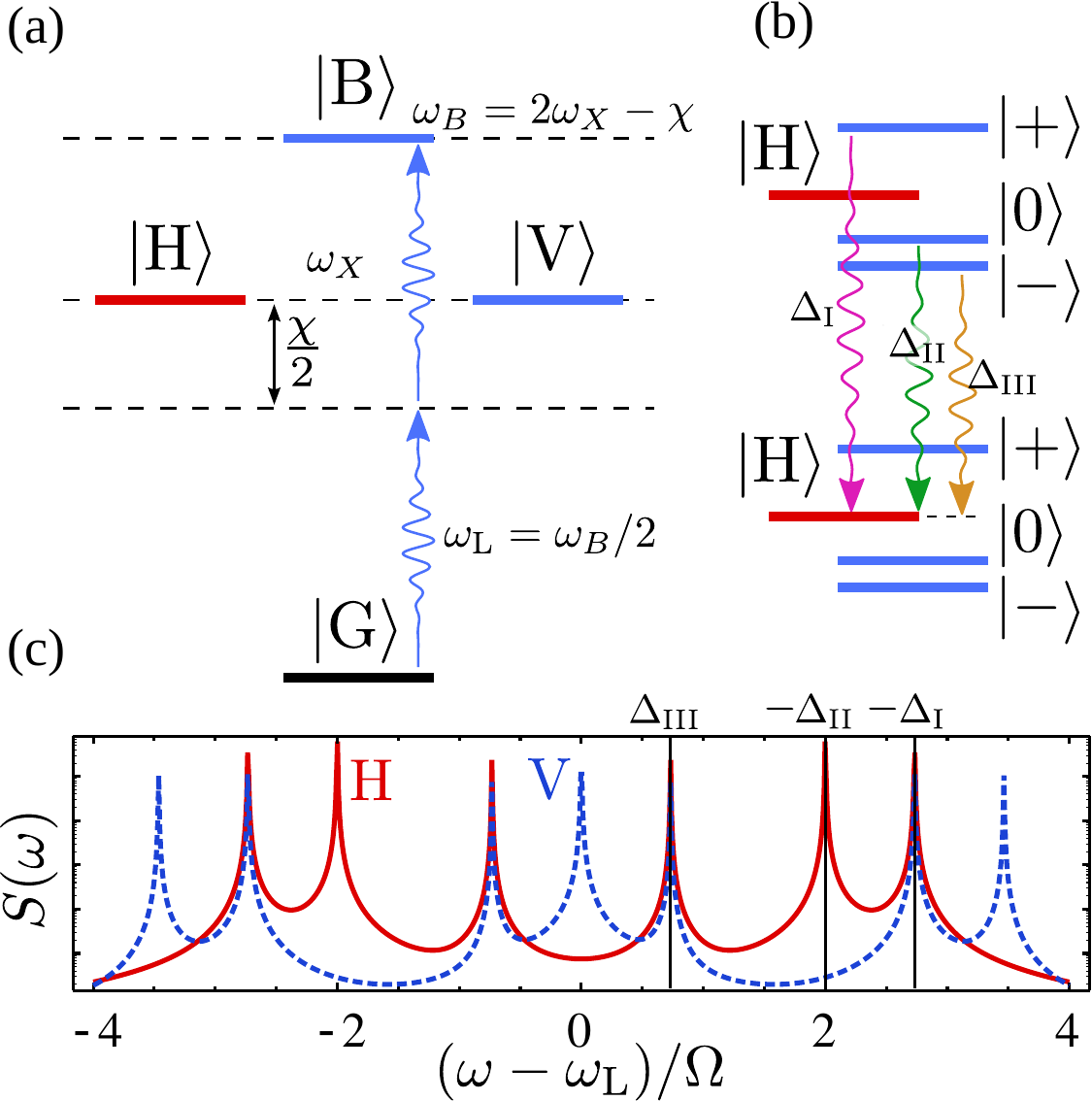}
\end{center}
\caption{(Color online) (a) Biexcitonic level system in the linear
  polarization basis (Horizontal-Vertical states). The two-photon
  laser excitation is represented with two curly blue arrows, at half
  the biexciton energy
  $\omega_\mathrm{L}=\omega_\mathrm{X}-\chi/2$. (b)~Dressed state
  picture at strong laser pumping, where the vertical polarization
  states (blue) transform into the new states $\ket{\pm}$,
  $\ket{0}\}$. The three possible horizontally polarised transitions
  from these states to the $\ket{\mathrm{H}}$, appear with curly
  arrows. (c)~Spectrum of emission in the two polarizations,
  horizontal (solid red) and vertical (dashed blue), and the three
  horizontally polarised transitions marked with vertical
  lines. Parameters: $\Omega=5\times10^2\gamma$, $\chi=2\times 10^3
  \gamma$ and $g=0$.}
\label{fig:1}
\end{figure}

The system under consideration is a semiconductor quantum dot with a
biexcitonic structure, as depicted in Fig.~\ref{fig:1}(a). The
Hamiltonian of this system is given by:
\begin{equation}
  \label{eq:FriDec5165349CET2014}
  H_\mathrm{X}\ = \omega_\mathrm{X}({\sigma_\uparrow}^\dagger \sigma_\uparrow + {\sigma_\downarrow}^\dagger \sigma_\downarrow)- \chi ({\sigma_\uparrow}^\dagger\sigma_\uparrow{\sigma_\downarrow}^\dagger\sigma_\downarrow),
\end{equation}
where $\{\sigma_\uparrow,\sigma_\downarrow\}$ are the annihilation
operators of the excitons with spin $\{\uparrow,\downarrow \}$,
$\omega_\mathrm{X}$ is the excitonic energy (we consider degenerate
excitons) and $\chi$ is the biexcitonic binding energy. The biexciton
frequency is, therefore,
$\omega_\mathrm{B}=2\omega_\mathrm{X}-\chi$. To separate the
four-level system into two different polarization cascades we change to the linear polarization basis:
\begin{equation}
  \label{eq:FriDec5165358CET2014}
  \ket{\mathrm{H}}=\frac{1}{\sqrt{2}}(\ket{\uparrow}+\ket{\downarrow}), \quad \ket{\mathrm{V}}=\frac{1}{\sqrt{2}}(\ket{\uparrow}-\ket{\downarrow})
\end{equation}
with the annihilation operators
\begin{equation}
  \label{eq:FriDec5165408CET2014}
  \sigma_\mathrm{H}=\frac{1}{\sqrt{2}}(\sigma_\uparrow + \sigma_\downarrow), \quad
  \sigma_\mathrm{V}=\frac{1}{\sqrt{2}}(\sigma_\uparrow - \sigma_\downarrow).
\end{equation}
These operators describe transitions from the biexciton to an
excitonic state or from an exciton to the ground state by emission of
photons with the corresponding horizontal or vertical polarization
(red and blue colors in Fig.~\ref{fig:1}). We will neglect the small
fine structure splitting in frequency that is usually found between
the two different excitonic states, since it has no impact in our
scheme and only complicates the algebra. It can be trivially added if
needed.

We implement a continuous resonant excitation of this level structure
that affects only one of the polarizations (chosen to be the vertical
one without loss of generalization). This is accounted for by a
coherent driving term in the Hamiltonian:
\begin{equation}
  H_\Omega = \Omega \, (\sigma_\mathrm{V}^\dagger e^{-i \omega_\mathrm{L} t} + \sigma_\mathrm{V} e^{i \omega_\mathrm{L} t})
\label{eq:HOmega}
\end{equation}
where $\Omega$ represents the intensity of the exciting laser. In
order to drive the biexciton state directly, the laser frequency is
set at the two-photon resonance,
$\omega_\mathrm{L}=\omega_\mathrm{B}/2$. This results in a two-photon
excitation (TPE) to the biexciton
level~\cite{flissikowski04a,stufler06a,jayakumar13a,muller14a,arxiv_huber15a}. On the
other hand, we gather and enhance the emission in the perpendicular
polarization (horizontal) through the coupling to a cavity mode with
the same linear polarization. This way, we completely separate in
polarization the excitation and emission channels and do not need to
worry about the elastically scattered light from the laser. The
coupling to the cavity mode is given by the Hamiltonian term:
\begin{equation}
\label{eq:FriDec5165426CET2014}
  H_\mathrm{C} = \omega_\mathrm{C} a^\dagger a + g(a^\dagger \sigma_\mathrm{H} + a\, \sigma_\mathrm{H}^\dagger)
\end{equation}
We write the total Hamiltonian in the rotating frame of the exciting
laser:
\begin{multline}
  H = H_0 + H_\Omega + H_\mathrm{C} =\\
  \Delta_\mathrm{X}(\shp\sh+\svp \sv)-\chi \shp \sh \svp \sv+\Omega(\svp+\sv)\\
  +\Delta_\mathrm{C} a^\dagger a + g(a^\dagger \sh+ a\shp)
\end{multline}
where $\Delta_\mathrm{X} = \omega_\mathrm{X}-\omega_\mathrm{L}$ and
$\Delta_\mathrm{C} = \omega_\mathrm{C} -\omega_\mathrm{L}$.  The
dynamics of the whole system is described by a density matrix
which follows the master equation:
\begin{equation}
\label{eq:FriDec5165434CET2014}
  \dot{\rho} = -i\left[H,\rho \right] + \frac{\kappa}{2}\mathcal{L}_a \rho+\frac{\gamma}{2} \sum_{\mathrm{X=H,V}}\left[\mathcal{L}_{\ket{\mathrm{X}}\bra{\mathrm{B}}}+\mathcal{L}_{\ket{\mathrm{G}}\bra{\mathrm{X}}} \right]\rho
\end{equation}
where we use the definition of the Lindblad term:
\begin{equation}
\label{eq:FriDec5165441CET2014}
\mathcal{L}_O \rho =2 O \rho O^\dagger - O^\dagger O \rho - \rho O^\dagger O \,,
\end{equation}
and the excitonic and cavity lifetimes are given by $\gamma$ and
$\kappa$ respectively. We study the steady state of the system defined
by $\dot{\rho} =0$.

Under TPE ($\Delta_\mathrm{X}=\chi/2$), the energy of the photons from the laser matches half the
biexciton energy, c.f.~Fig.~\ref{fig:1}(a). To understand the spectral
features of the system before coupling it to the cavity ($g=0$), we
derive a dressed state picture for the biexciton~\cite{cohentannoudji77a}. The starting point is the set of
bare states with $n$ excitations,
$\left\{\ket{\mathrm{G}}\ket{n},\ket{\mathrm{V}}\ket{n-1},\ket{\mathrm{H}}\ket{n-1},\ket{\mathrm{B}}\ket{n-2}
\right\}$, where $\ket{n}$ describes the state of the driving field with $n$
photons. Since the laser is polarized in the vertical direction, the
state $\ket{\mathrm{H}}$ is not dressed by it, while the rest of the excitonic
states are. The new eigenstates are obtained by diagonalising the
coupling Hamiltonian~$H_\Omega$ (in the rotating frame of the laser)
in the reduced basis
$\left\{\ket{\mathrm{G}}\ket{n},\ket{\mathrm{V}}\ket{n-1},\ket{\mathrm{B}}\ket{n-2}
\right\}$, that is, the matrix:
\begin{equation}
\label{eq:FriDec5165448CET2014}
  H_{\mathrm{TPE}} = \begin{pmatrix}
    0&\Omega&0\\
    \Omega&\chi/2&\Omega\\
    0&\Omega&0
  \end{pmatrix}\,.
\end{equation}
We do not include dissipation in this procedure since we consider it
small as compared to $\Omega$. This gives rise to the three new
eigenvectors $\{ \ket{+},\ket{0},\ket{-}\}$ in each rung with the
corresponding eigenenergies:
\begin{subequations}
\label{eq:FriDec5165455CET2014}
\begin{align}
  &\Delta_+=\frac{1}{4}\left(\sqrt{\chi^2 + 32 \Omega^2}+\chi \right)\,,\\
  &\Delta_0 =0\,,\\
  &\Delta_-=-\frac{1}{4}\left(\sqrt{\chi^2 + 32 \Omega^2}-\chi \right)\,,
\end{align}
\end{subequations}
where the eigenvectors, dropping the photonic component from the
notation, are given by $\ket{+} \propto
\ket{\mathrm{G}}+\Delta_+/\Omega \ket{\mathrm{V}} + \ket{\mathrm{B}}$,
$\ket{0}=(\ket{\mathrm{B}}-\ket{\mathrm{V}})/\sqrt{2}$ and $\ket{-}
\propto \ket{\mathrm{G}}+\Delta_-/\Omega \ket{\mathrm{V}} +
\ket{\mathrm{B}}$.  Figure~\ref{fig:1}(b) depicts two successive rungs
of excitation, including the state $\ket{\mathrm{H}}$ which, as we said,
remains bare.

\section{Single photon transition and spectrum}

The spectrum of emission of the system in each polarization in the
steady state, $S_\mathrm{X}(\omega)$, with X=H, V, is defined as
$S_\mathrm{X}(\omega)=\Re
\int_{0}^{\infty}\mean{\sigma_\mathrm{X}^\dagger(0)\sigma_\mathrm{X}(\tau)}e^{i\omega
  \tau }d\tau$. Both polarizations are plotted in Fig.~\ref{fig:1}(c)
for comparison. The number of peaks appearing and their positions can
be explained in each polarization X by the allowed single­ photon
transitions under the operator $\sigma_\mathrm{X}$.  In the case of H
polarization, only transitions between $\ket{\mathrm{H}}$ and the
dressed states $i=+,0,-$ are allowed: $|\bra{\mathrm{H}}\sh
\ket{i}|^2\neq 0$ or $|\bra{i}\sh \ket{\mathrm{H}}|^2\neq 0$. The
transition $\ket{\mathrm{H}}\rightarrow\ket{\mathrm{H}}$ or between
dressed states $\ket{i}\rightarrow \ket{j}$ are forbidden in H
polarization, since $|\bra{\mathrm{H}}\sh \ket{\mathrm{H}}|= 0$ and
$|\bra{i}\sh\ket{j}|=0$ for all $i,j=+,0,-$. The three possible
transitions that can take place from the dressed states to
$\ket{\mathrm{H}}$, occur respectively at the following detunings from
the laser (see Fig.~\ref{fig:1}(b)):
\begin{subequations}
\label{eq:FriDec5165502CET2014}
\begin{align}
  \ket{+}\rightarrow \ket{\mathrm{H}}&: \quad \Delta_\mathrm{I}  =\frac{1}{4}\left(\sqrt{\chi^2 + 32 \Omega^2} -\chi\right)\,,\\
  \ket{0}\rightarrow \ket{\mathrm{H}}&: \quad \Delta_\mathrm{II}=-\chi/2\,,\\
  \ket{-}\rightarrow \ket{\mathrm{H}}&: \quad \Delta_\mathrm{III}=-\frac{1}{4}\left(\sqrt{\chi^2 + 32 \Omega^2}+\chi \right)\,.
\end{align}
\end{subequations}
The other three possible H-polarised transitions take place from
$\ket{\mathrm{H}}$ to the dressed states, at opposite detunings
$-\Delta_\mathrm{I}$, $-\Delta_\mathrm{II}$ and
$-\Delta_\mathrm{III}$.  Remarkably, $S_\mathrm{H}(\omega)$ does not
present any resonance at the laser energy.

On the other hand, the spectrum in V polarization,
$S_\mathrm{V}(\omega)$, plotted with a dashed blue line in
Fig.~\ref{fig:1}(c), contains seven peaks corresponding to the nine
possible transitions between dressed states, $\ket{i}\rightarrow
\ket{j}$, with those three between the same dressed states,
$\ket{i}\rightarrow \ket{i}$, degenerate in energy at
$\omega_\mathrm{L}$.

\section{Two-photon transitions and spectrum}

\begin{figure*}[t]
\begin{center}
\includegraphics[width=\linewidth]{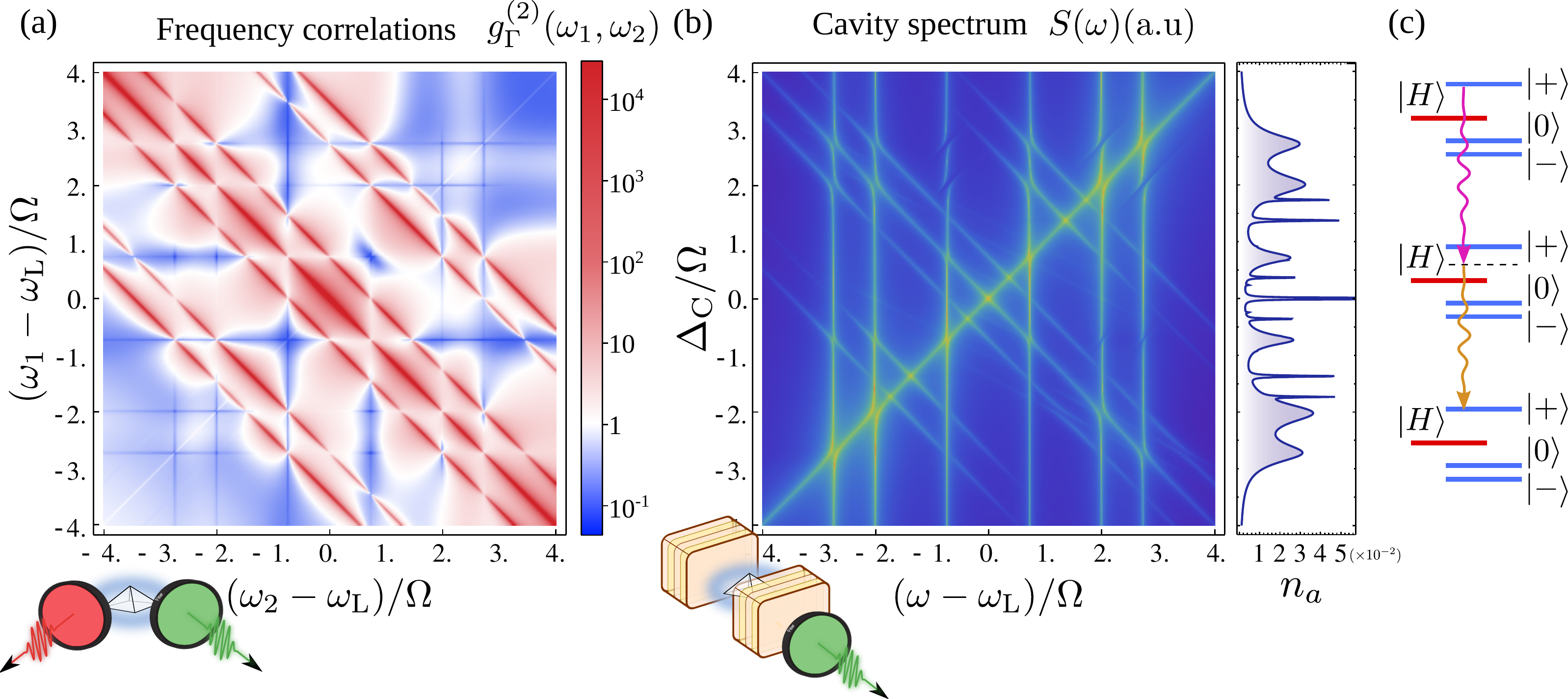}
\end{center}
\vspace{-10pt}
\caption{(Color online) (a) Two-photon spectrum in H polarization for
  the TPE. In blue, sub-Poissonian statistics (antibunching), in red,
  super-Poissonian statistics (bunching) and in white, Poissonian
  statistics (uncorrelated). Parameters: $\chi=4\times 10^3\gamma$,
  $\Omega=10^3 \gamma$, $\Gamma=10\gamma$ and $g=0$. (b) Cavity
  spectrum of emission as a function of the cavity frequency
  $\omega_a$ in the strong coupling regime, $g=10^2\gamma$,
  $\kappa=10\gamma$. The plot on the right hand side shows the
  integrated signal, i.e., the cavity population $n_a$. (c) Example of the
  two-photon transition $\ket{+}\rightarrow\rightarrow\ket{+}$ in the
  H polarization.}
\label{fig:2}
\end{figure*}

\begin{figure}[b]
\begin{center}
\includegraphics[width=0.9\linewidth]{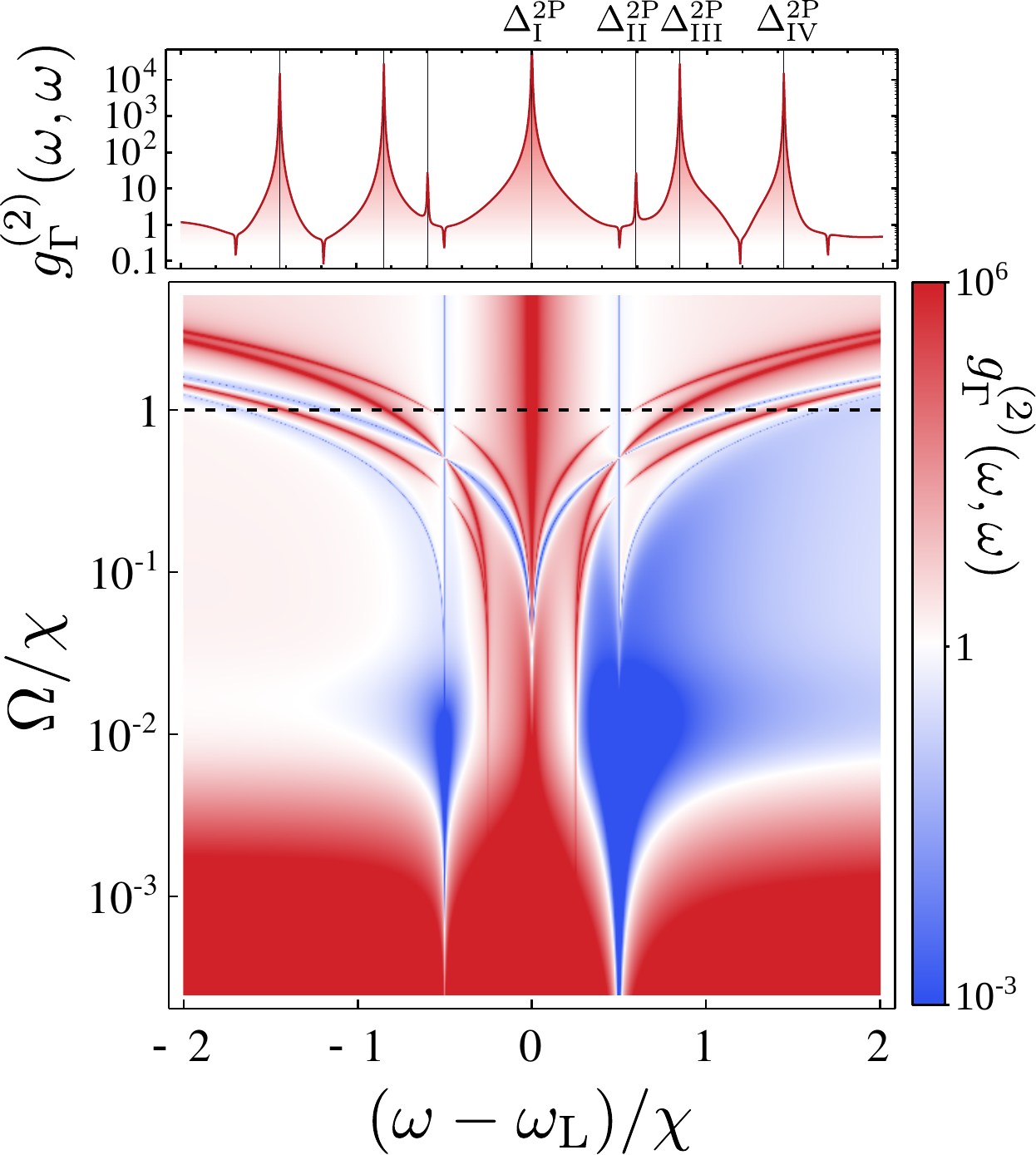}
\end{center}
\caption{(Color online) Two-photon spectrum with
  $\omega=\omega_1=\omega_2$ (the diagonal) of the biexciton system as
  a function of the driving field intensity $\Omega$. Top panel shows a cut along the dashed line in the bottom panel. Blue colors in the map represent
  sub-Poissonian statistics (antibunching), red, super-Poissonian
  statistics (bunching) and white, Poissonian statistics
  (uncorrelated). The physics changes from that of the biexciton
  spontaneous emission regime (bottom part), with a single leapfrog
  peak, to that of the dressed biexciton system (top part), with seven
  leapfrog peaks. Parameters: $\chi=4\times 10^3\gamma$,
  $\Gamma=10\gamma$ and $g=0$.}
\label{fig:3}
\end{figure}

The next step in the characterization of the system is the calculation
of the frequency-resolved second order correlation function or
two-photon spectrum,
$g^{(2)}_\Gamma(\omega_1,\omega_2)$~\cite{delvalle12a,gonzaleztudela13a},
that conveys how likely is to detect two photons with frequencies
$\omega_1$, $\omega_1$ simultaneously. For that purpose we use a
recently developed technique~\cite{delvalle12a} that makes the
calculation of this quantity, previously very involved,
computationally accessible, based on the inclusion of the detectors in
the system dynamics.  The parameter $\Gamma$ is the inverse response
time of the detector. It provides the frequency window in which
photons are detected around $\omega_1$, $\omega_2$. We fix it to an
intermediate value $\Gamma=10\gamma$, so that the detectors can
resolve full spectral peaks (with width of the order of $\gamma$)
without resulting in superimposed signals, $\gamma<\Gamma\ll \Omega$.

Figure~\ref{fig:2}(a) shows the H-polarized two-photon spectrum from
the light emitted by the dressed biexciton system. This map features
seven antidiagonal red lines of super Poissonian correlations
with~$g^{(2)}_\Gamma\gg1$ (hyperbunching) that correspond to a family
of virtual two-photon processes that go from one state in a rung to
another state two rungs below, jumping over any states from the rung
in between (whence the denomination of \emph{leapfrog} processes). It
was recently demonstrated~\cite{sanchezmunoz14b} that this virtual
character provides such strong quantum correlations that photon pairs
can violate classical inequalities such as the Cauchy-Schwarz
inequality. Whenever any two of the frequencies involved correspond to
transitions between real states, these correlations change character
and the violation of Cauchy-Schwarz inequalities is spoiled. This can
be seen in Fig.~\ref{fig:2}(a) as a piercing in the bunching lines
whenever they intersect the vertical or horizontal ones, appearing at
$\omega_{1,2}-\omega_\mathrm{L}=\pm\Delta_\mathrm{I}$,
$\pm\Delta_\mathrm{II}$, $\pm\Delta_\mathrm{III}$.

Since the leapfrog lines originate from two-photon transitions, we
can understand them in terms of the two-photon operator
$\sigma_\mathrm{H}\sigma_\mathrm{H}$. Transitions starting or ending
at $\ket{\mathrm{H}}$ are not allowed, since
$|\bra{\mathrm{H}}\sh\sh\ket{i}|=0$ and
$|\bra{i}\sh\sh\ket{\mathrm{H}}|=0$. All other nine two-photon
transitions, %$\ket{i}\rightrightarrows\ket{j}$, occur, since
$\ket{i}\rightarrow\rightarrow\ket{j}$, occur, since
$|\bra{j}\sh\sh\ket{i}|\neq 0$ for all $i,j=+,0,-$, and give rise to
seven lines which follow the general equation
$\omega_1+\omega_2-2\omega_\mathrm{L}=\Delta^\mathrm{2P}$ with:
\begin{subequations}
\begin{align}
&\ket{i}\rightarrow\rightarrow\ket{i}\quad \mathrm{with}\quad i=+,0,-:\quad  \Delta_\mathrm{I}^\mathrm{2P} = 0 \,,\label{eq:lfI}\\
&\ket{+}\rightarrow\rightarrow\ket{0}:\quad  \Delta_\mathrm{II}^\mathrm{2P}=
\frac{1}{8}\left(\sqrt{\chi^2 + 32 \Omega^2} +\chi\right) \,,\label{eq:lfII}\\
&\ket{0}\rightarrow\rightarrow\ket{-}:\quad \Delta_\mathrm{III}^\mathrm{2P}=
\frac{1}{8}\left(\sqrt{\chi^2 + 32 \Omega^2} -\chi\right)\,, \label{eq:lfIII}\\
&\ket{+}\rightarrow\rightarrow\ket{-}:\quad  \Delta_\mathrm{IV}^\mathrm{2P}=\frac{1}{4}\sqrt{\chi^2 + 32 \Omega^2} \,.\label{eq:lfIV}
\end{align}
\label{eq:leapfrogs}
\end{subequations}
The remaining three lines are described by inverting the order of the
three last transitions and changing the sign of the corresponding
$\Delta^\mathrm{2P}$. Figure~\ref{fig:2}(c) shows an example of a
two-photon transition, $\ket{+}\rightarrow\rightarrow\ket{+}$.

In Fig.~\ref{fig:3}, we have another view of these leapfrog resonances,
selecting the diagonal of the two-photon spectrum in
Figure~\ref{fig:2}(a), that is, for $\omega=\omega_1=\omega_2$. The
leapfrog processes appear as seven lines around $\Omega/\chi\approx
10^{-1}$ and spread as $\Omega$ is increased. The blue lines
correspond to the single-photon resonances that are also apparent in
the spectrum of emission, c.f.~Fig.~\ref{fig:1}(c). Reducing $\Omega$
below the dissipation levels (bottom part of the plot), the system
experiences a transition into the spontaneous emission regime where
there is no dressing of the levels and the spectral structures are
much simpler: only two peaks for the spectrum of emission and a single
leapfrog peak at $\omega=\omega_\mathrm{L}$ in the two-photon
spectrum. This regime was extensively investigated by one of the
authors under incoherent excitation~\cite{delvalle13a}. In the present
work, where it appears as the low pumping limit, it will be used only
for comparison with the high pumping regime.

\vspace{15pt}
\section{Purcell enhancement of two-photon transitions by a cavity mode}

These virtual leapfrog transitions can be made real by coupling the
system to a cavity (we switch on $g\neq 0$) in resonance with at least
one of the two frequencies involved. If the coupling is strong as
compared to the cavity dissipation, $\kappa$, the two-photon emission
can be Purcell-enchanced. We can observe this in the cavity spectrum
of emission, given by $S_a(\omega) =\Re
\int_{0}^{\infty}\mean{a^\dagger(0)a(\tau)}e^{i\omega \tau }d\tau$ and
plotted in Fig.~\ref{fig:2}(b). Because of the strong correlations
between the two frequencies, the cavity Purcell-enhancement of one of
the two photons of a bunching line triggers the emission of the second
photon, even if this one is not in resonance with the cavity. This
phenomenon leaves traces in the spectrum that help reconstruct the
bunching lines when the spectrum is plotted as a function of the
cavity frequency. In this sense, the cavity is acting as one of the
filters necessary to perform frequency correlations.

As we have shown with coworkers in a recent
work~\cite{sanchezmunoz14a}, one can obtain useful two-photon emission
by using this approach to Purcell-enhance two photons of the same
frequency. This is evidenced by sharp peaks in the cavity population
whenever it crosses one of the two-photon resonances
(Eqs.~\eqref{eq:leapfrogs} with $\omega_a=\omega_1=\omega_2$), as can
be seen in the right panel of Fig.~\ref{fig:2}(b). The single photon
resonances appear as broad peaks and are detrimental for the
two-photon emission. Therefore, the best candidates for pure
two-photon emission are those leapfrogs far in energy from other
processes, that is, the sharp peaks with small overlap with the
(one-photon) broad ones and that are further from other (two-photon)
sharp ones. Logically, it is also desirable that they are intense. The
central peak, labeled I, at $\omega_a=\omega_\mathrm{L}$ is the best
candidate since it is the most isolated one and is degenerate, with
contributions from three different leapfrog processes. As we will
discuss, this has consequences on the statistics of the emitted pairs.

\begin{figure}[b!]
\begin{center}
\includegraphics[width=0.8\linewidth]{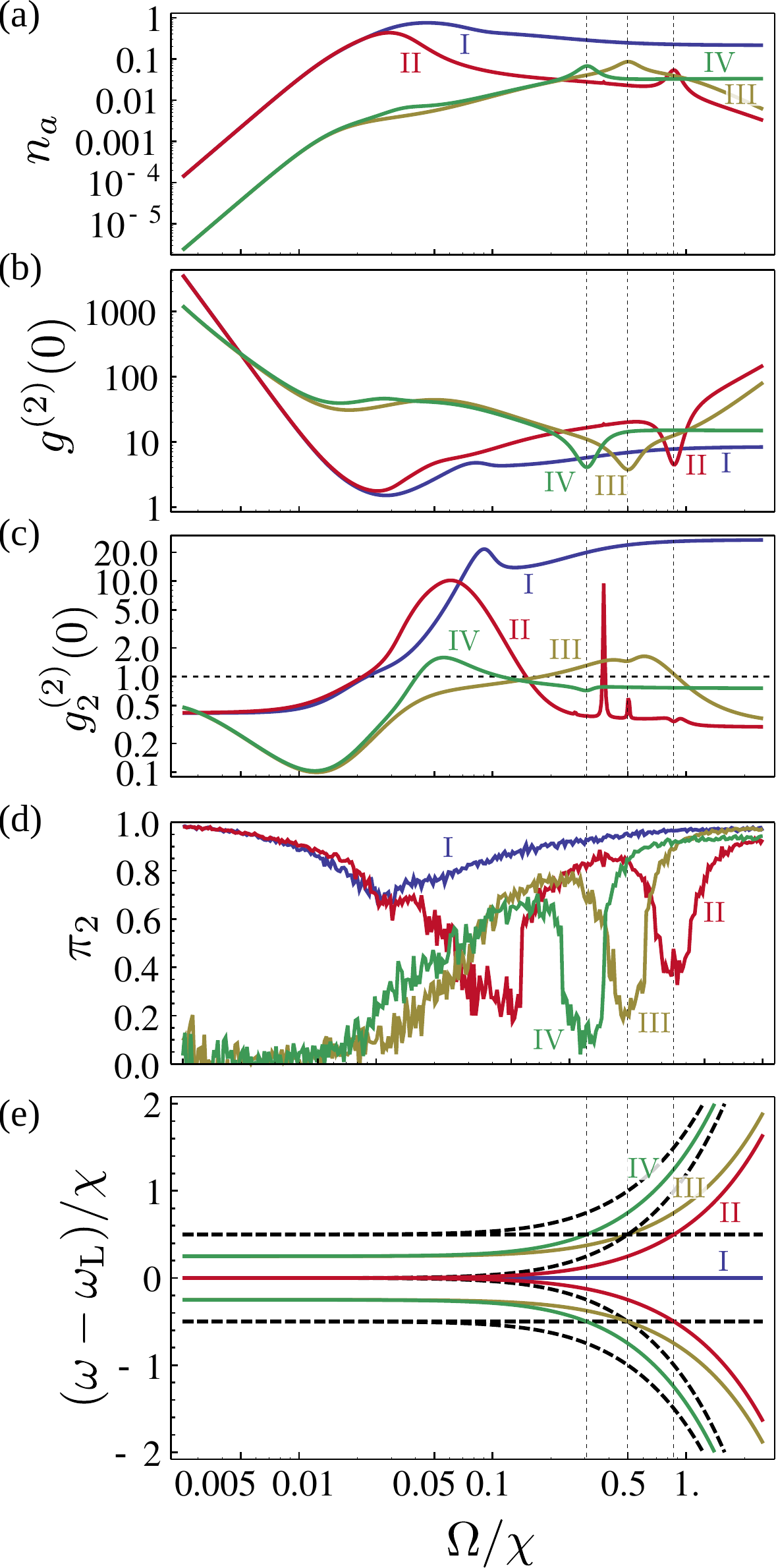}
\end{center}
\caption{(Color online) Steady state observables as a function of the pumping intensity $\Omega$ for the parameters: $\chi=4\times 10^3\gamma$,
  $g=10^2 \gamma$, $\kappa = 10\gamma$, $\Delta_\mathrm{X}= \chi/2$ and $\Delta_\mathrm{C} = \Delta_{\mathrm{I}}^{\mathrm{2P}}$ (blue), $\Delta_{\mathrm{II}}^{\mathrm{2P}}$ (red), $\Delta_{\mathrm{III}}^{\mathrm{2P}}$ (yellow) and $\Delta_{\mathrm{IV}}^{\mathrm{2P}}$ (green). The gridlines mark the three points where leapfrog processes intersect with real transitions---dashed lines in (e)---, which spoils the purity. }
\label{fig:4}
\end{figure}

An accurate quantity to determine the quality of a two-photon
resonance for two-photon emission is the \emph{purity},
$\pi_2$~\cite{sanchezmunoz14a}, defined as the fraction of photons
emitted in pairs from the total emission (including single
photons). Note that the purity being a probability, it is, unlike
$g^{(2)}$, bounded: $0\le\pi_2\le1$. Its definition is based on the
fact that the photon counting distribution of a perfect two-photon
emitter shows a suppressed probability of counting an odd number of
photons. The details of its definition and computation can be found in
the Supplemental Material.  In order to compute it, we simulate the
actual emission of the system in the steady state via a quantum
Monte-Carlo method~\cite{sanchezmunoz14a}. The result is plotted in
Fig.~\ref{fig:4}(d) for a cavity on resonance with each of the
leapfrog peaks in the two-photon spectrum: I, II, III and IV, whose
positions shift with $\Omega$ as plotted in panel~(e). The
corresponding cavity population~$n_a=\mean{a^\dagger a}$, second order
correlation function $g^{(2)}(0)=\mean{{a^\dagger}^2 a^2}/n_a^2$ and
two-photon second order correlation function~\cite{sanchezmunoz14a}
$g^{(2)}_2(0)=\mean{{a^\dagger}^4 a^4}/\mean{{a^\dagger}^2 a^2}^2$
appear in (a), (b) and (c) respectively. The latter quantity takes the
meaning of a standard \emph{second order correlation function for
  photon pairs} when these pairs dominate the emission ($\pi_2 \approx
1$).

On the low driving regime, we can see that resonances I and II
converge to the same point at $\omega = \omega_\mathrm L$ and show
very high purity: this is the usual regime of two-photon emission in
the (undressed) biexciton, that has been studied extensively
before~\cite{delvalle11d,delvalle12b}. Note, however, that this high
purity comes at the expense of the amount of signal (low $n_a$). As
Fig.~\ref{fig:4} shows, this signal can be enhanced by orders of
magnitude if we increase the pumping intensity $\Omega$ in order to
bring the biexciton to the dressed regime. In this regime, all the
resonances start being resolved and the four of them present a sizable
purity. In the case of resonances II, III and IV, the purity goes down
whenever they cross a single-photon resonance (dashed, vertical lines
in Fig~\ref{fig:4}).  At very high intensity, $\Omega>\chi$, all of
them reach almost 100\% of pair emission.

In this limit of high pumping, we observe a bunching behaviour
$g^{(2)}(0)>1$ for all the leapfrog resonances, which is an expected
result for two-photon emission. The statistics of the photons,
however, hides a non-classical behaviour if one regards the pairs as
the basic entity of emission and consider the pair-pair coincidences
as described by the $g^{(2)}_2(0)$. In this case, we obtain
\emph{antibunched photon pairs} in resonances II, III and IV and
bunched photon pairs in resonance I.  This is not the only difference
between resonance I (blue) and the others. The position of this
resonance is independent of the pumping intensity, and the emission at
this frequency is order of magnitudes more intense than at the other
resonances. These differences are explained by the fact that three
different transitions contribute to the photon pair emission at line
I, and that the starting and ending state of the transition are always
the same, as can be seen in Eq.~\eqref{eq:lfI}. Because of this, no
reloading time to go back to the initial state is needed, which is the
origin of the antibunching on all the other cases. All these features
are a sample of the rich set of physical regimes that can be explored
when the proposed method of multi-photon Purcell enhancement in
dressed states systems is applied in non-trivial configurations.

\section{Emission of entangled photons}
\begin{figure}[b!]
\begin{center}
\includegraphics[width=1\linewidth]{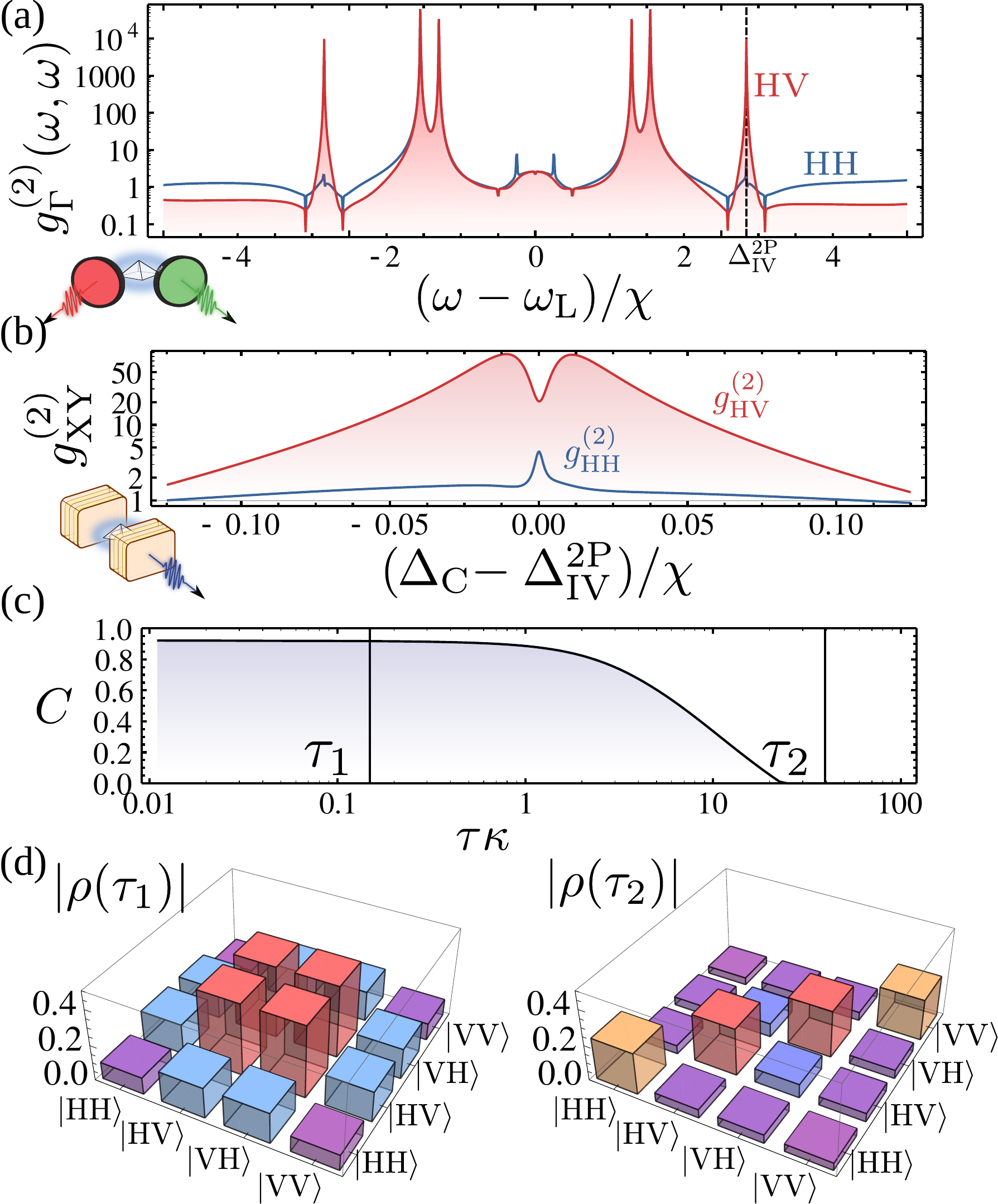}
\end{center}
\caption{(Color online) (a) Two photon spectrum for $\omega=\omega_1=\omega_2$ (diagonal) of the dressed biexciton for photons with opposed polarization (red) and same polarization (blue). For photons emitted at frequency $\Delta_\mathrm{IV}^\mathrm{2P}$, the cross correlation between polarizations is higher than the autocorrelations. (b) Autocorrelations and cross-correlations between two cavity modes with orthogonal polarizations coupled to the dressed biexciton as a function of their energy around $\Delta_\mathrm{IV}^\mathrm{2P}$. (c) Concurrence of the emitted photon-pair state as a function of the total measurement time. (d) Density matrix of the emitted state for two different total measurement times denoted as $\tau_1$ and $\tau_2$. Parameters: $\Omega = 8\times 10^3 \gamma$, all the rest same as in Fig.~\ref{fig:4}.   }
\label{fig:5-2}
\end{figure}

Many practical applications in quantum computing and quantum
communication require emission of entangled photon
pairs~\cite{pan12a,jennewein00a,naik00a,bouwmeester97a,marcikic03a,pan98a,simon07a,troiani14a}. So
far, we have only considered the case in which the emission was
filtered by a single cavity with a fixed polarization. Therefore,
changing the frequency of the cavity corresponds to moving along the
diagonal of Fig.~\ref{fig:2}(a), and all photons extracted by the
cavity will tend to be indistinguishable. However, the results for the
spectrum of the cavity emission depicted in Fig.~\ref{fig:2}(b) show
that correlated photons of different frequencies can be Purcell
enhanced with a single cavity: therefore, two cavities in resonance
with two different, correlated frequencies---showing bunching in the
map of Fig.~\eqref{fig:2}(a)---will be expected to show strongly
correlated emission. In our case, another interesting possibility
provided by the biexcitonic structure is to work with two degenerate
polarizations of a single cavity, described by the bosonic
annihilation operators $a_\mathrm{H}$ and $a_\mathrm{V}$. Two-photon
emission then takes place in a reduced Hilbert space of polarizations
$\{\ket{\mathrm{HH}},\ket{\mathrm{HV}} ,\ket{\mathrm{VH}}
,\ket{\mathrm{VV}} \}$ of photons with the same energy. We will now
show how the mechanism of two-photon emission described so far can
also yield emission of entangled photons of the kind
$\ket{\psi}=(\ket{\mathrm{HV}}+\ket{\mathrm{VH}})/\sqrt{2}$. The
problem is theoretically described in the same way as before, but now
with a different coupling Hamiltonian:
\begin{eqnarray}
\label{eq:HcouplingPol}
  H_\mathrm{C} &=& \omega_\mathrm{C}(a_\mathrm{H}^\dagger a_\mathrm{H} +a_\mathrm{V}^\dagger a_\mathrm{V})\nonumber \\
  &+& g(a_\mathrm{H}^\dagger \sigma_\mathrm{H} + a_\mathrm{H}\, \sigma_\mathrm{H}^\dagger)+ g(a_\mathrm{V}^\dagger \sigma_\mathrm{V} + a_\mathrm{V}\, \sigma_\mathrm{V}^\dagger)
\end{eqnarray}
The vertically polarized driving term \eqref{eq:HOmega}  used above leads to different probability of emission in horizontal or vertical polarization. Since we now want equal probability, we use a circularly-­polarized driving laser:
\begin{equation}
  H_\Omega = \Omega \, (\sigma_\mathrm{ \circlearrowright}^\dagger e^{-i \omega_\mathrm{L} t} + \sigma_\mathrm{ \circlearrowright} e^{i \omega_\mathrm{L} t})
\end{equation}
with $\sigma_\mathrm{ \circlearrowright}=(\sigma_\mathrm{H} +i
\sigma_\mathrm{V})/\sqrt{2}$. We will not discuss in detail the
possible single and two-photon transitions that arise in this model,
since the physics and derivation are similar to the above
material. However, it is now of interest to analyze the
frequency-resolved, cross polarized second order correlation function
$g^{(2)}_{\Gamma,\mathrm{HV}}(\omega_1,\omega_2)$ of the dressed
biexcitonic system alone ($g=0$), which is a cross-correlation
function between photons emitted at frequency $\omega_1$ and
polarization H and photons emitted at frequencies $\omega_2$ and
polarization V. This correlation function can be compared with the
frequency-resolved correlation functions for a fixed polarization
that we have been considering so far, that we now term
$g^{(2)}_{\Gamma,\mathrm{HH}}(\omega_1,\omega_2)$. Due to the circular
polarized pumping, the system is symmetric under the exchange
$\mathrm{H}\leftrightarrow \mathrm{V}$, so
$g^{(2)}_{\Gamma,\mathrm{HH}}(\omega_1,\omega_2)=g^{(2)}_{\Gamma,\mathrm{VV}}(\omega_1,\omega_2)$
and
$g^{(2)}_{\Gamma,\mathrm{HV}}(\omega_1,\omega_2)=g^{(2)}_{\Gamma,\mathrm{VH}}(\omega_1,\omega_2)$.

Figure~\ref{fig:5-2}(a) shows the comparison between these two
quantities, with both photons having the same frequency ($\omega_1 =
\omega_2 = \omega$), equivalent to Fig.~\ref{fig:3}, which was for the
case of linearly polarized pumping. We observe that photons emitted at
a frequency $\omega =
\omega_\mathrm{L}+\Delta_{\mathrm{IV}}^\mathrm{2P}$ present strong
cross-polarized correlations, clearly higher than the autocorrelation
for each polarization. This is a non-classical result that violates
the Cauchy-Schwarz inequality~\cite{reid86a,sanchezmunoz14b}. When the
two degenerate modes described by $a_\mathrm{H}$ and $a_\mathrm{V}$
are tuned to $\omega_\mathrm{L}+\Delta_{\mathrm{IV}}^\mathrm{2P}$ and
coupled to the system as described in Eq.~\eqref{eq:HcouplingPol}, their
cross correlation $g^{(2)}_{\mathrm{HV}}=\langle a^\dagger_\mathrm H
a^\dagger_\mathrm V a_\mathrm H a_\mathrm V \rangle/(\langle
a^\dagger_\mathrm H a_\mathrm H \rangle \langle a^\dagger_\mathrm V
a_\mathrm V \rangle)$ and autocorrelation
$g^{(2)}_{\mathrm{HH}}=\langle a^\dagger_\mathrm H a^\dagger_\mathrm H
a_\mathrm H a_\mathrm H \rangle/\langle a^\dagger_\mathrm H a_\mathrm
H \rangle^2$ replicate the trend just described for the
frequency-resolved second order correlations of the dressed biexciton
alone. This can be seen in Figure~\ref{fig:5-2}(b), where the auto and
cross-correlations of the two cavity modes are shown for a range of
cavity frequencies around
$\omega_\mathrm{L}+\Delta_{\mathrm{IV}}^\mathrm{2P}$.

Quantum tomography~\cite{james01a,dousse10a,troiani06b,delvalle13a}
allows us to reconstruct the density matrix of the emitted photon
pairs in the basis $\{\ket{\mathrm{HH}},\ket{\mathrm{HV}}
,\ket{\mathrm{VH}} ,\ket{\mathrm{VV}} \}$ from second order
correlation functions corresponding to photon coincidence
measurements. We define the (unnormalized) density matrix:
\begin{equation}
\theta_\mathrm{AB,CD}(\tau)=\int_0^\tau \langle a_\mathrm{A}^\dagger(0) a_\mathrm{B}^\dagger(\tau') a_\mathrm{D}(\tau')  a_\mathrm{C}(0)\rangle d\tau'
\end{equation}
with $\mathrm{A,B,C,D}\in \{\mathrm{H,V}\}$, where two-time
correlation functions are calculated from the steady state of the
system using the quantum regression
theorem~\cite{scully_book02a}. Therefore, $\tau$ corresponds to the
time of measurement that begins with the emission of the first photon, and for each $\tau$ we define the normalized density matrix $\tilde{\theta}(\tau)=\theta(\tau)/\Tr[\theta(\tau)]$. This analysis reveals, for short measurement times, a highly
pure density matrix, $\mathrm{Tr}[\tilde\theta^2]\approx 0.92$ consisting of
the entangled Bell state
$\ket{\psi}=(\ket{\mathrm{HV}}+\ket{\mathrm{VH}})/\sqrt{2}$ with
fidelity $F\approx 0.9$, shown in Fig.~\ref{fig:5-2}(d). Beyond a
certain time of measurement $1/\kappa$,
the density matrix loses purity due to the contributions from
subsequent emissions. The degree of entanglement of this emitted
bipartite state can be quantified by the concurrence
$C$~\footnote{This quantity is defined as $C =
\max\{0,\sqrt{\lambda_1}-\sqrt{\lambda_2}-\sqrt{\lambda_3}
-\sqrt{\lambda_4} \} $, where $\{\lambda_1,
\lambda_2,\lambda_3,\lambda_4 \}$ are the eigenvalues in decreasing
order of the matrix $\rho T \rho^* T$, with $T$ a diagonal matrix
with diagonal $\{-1,1,1,-1 \}.$}, that in the case of pure states ranges from 0 (separable states)
to 1 (maximally entangled states)~\cite{wootters98a}. In our case, the concurrence takes a value $C \approx 0.92$ for short measurement times. However, one must bear in mind that the maximum concurrence for a mixed state is lower than one~\cite{wei03} and $\tilde\theta$ is a mixed state with linear entropy $S_L(\theta)=4/3[1-\Tr(\theta ^2)] \approx 0.11$, which brings this value of $C$ closer to that of a maximally entangled mixed state.
Not only is this result interesting by itself, but, being
just a particular example, it also suggests that photons entangled not
only in the polarization, but also in the energy degree of
freedom---which in this case has been chosen equal for both photons
for simplicity---could be obtained.

\section{Conclusions}

We have shown how Purcell enhancement of multi-photon resonances in
the dressed ladder of a strongly driven biexciton can yield regimes of
bright continuous two-photon emission.  Thanks to the strong driving,
the emission of photon pairs occurs at a much higher rate than it
would in the approach that combines standard TPE (without dressing the
system) and two-photon Purcell enhancement. The richness of the
dressed biexcitonic structure allows to reach different two photon
regimes like antibunched two-photon emission or entangled photon
pairs. These results suggest that the fundamental ideas behind this
particular proposal are susceptible to be applied in a variety of
platforms.

\section{Acknowledgments}

We acknowledge the IEF project SQUIRREL (623708), the Spanish MINECO
(MAT2011-22997, MAT2014-53119-C2-1-R, FPI \& RyC programs) and the ERC
POLAFLOW.

\bibliographystyle{naturemag}
\bibliography{Sci,books,arXiv}

% Supplemental material
\setcounter{equation}{0}
\setcounter{figure}{0}
\setcounter{table}{0}
\setcounter{page}{1}
\renewcommand{\theequation}{S\arabic{equation}}
\renewcommand{\thefigure}{S\arabic{figure}}
\renewcommand{\bibnumfmt}[1]{[S#1]}
\renewcommand{\citenumfont}[1]{S#1}
\pagenumbering{roman}

\vfill\eject

\onecolumngrid{
\begin{center}
\textbf{\large Enhanced two-photon emission from a dressed biexciton.\\ Supplemental Material}
\end{center}
}

In this Supplemental Material we define the \emph{purity} $\pi$, a magnitude that quantifies the percentage of photons emitted in pairs from the total emission. This is therefore a bounded quantity $\pi \in [0,1]$, and its definition is based on the observation that a perfect two-photon emitter will never emit an odd number of photons. We start the discussion by describing the probability distribution of the sum of two random variables in terms of generating functions. In our case, these two random variables correspond to the number of photons emitted by two different processes: one emits photons one by one, and the other, in pairs. The entirety of the light emitted is assumed to be the sum of these two processes, and we will show how this assumption does describes very accurately the photon counting distributions of the light emitted by the system.

\section{Sum of two random processes}
For a given discrete random process $X$, we can define the generating function $\Pi_X=\mean{s^X}$, in such a way that the probability of obtaining $X$ is given by $P(X)=\frac{1}{n!}\partial^{(x)}/\partial s^x  \Pi_X |_{s=0}$. When we combine different random processes, the generating function is given by the product of the two original ones: $\Pi_{X_1+X_2}=\Pi_{X_1} \Pi_{X_2}$. In the case of a coherent random process, the generating function is given by $\Pi_\lambda = e^{-\lambda(1-s)}$. As we will see, this exponential form will be convenient. For the purpose of obtaining a general expression for the $n$-th derivative needed to compute $P(X)$ from the generating function, we will use Fa\`a di Bruno's formula for the generalized chain rule expressed in term of Bell polynomials $B_{n,k}(x_1, \cdots,x_{n-k+1})$:
\begin{equation}
\frac{d^n}{dx^n}f(g(x)) = \sum_{k=1}^n f^{(k)}(g(x)) B_{n,k}\left(g'(x),g''(x),\cdots, g^{(n-k+1)}(x) \right).
\label{eq:brunos}
\end{equation}
If $f$, as is the case for the coherent process, is just the exponential function, the $n$-th derivative of $f$ appearing in \eqref{eq:brunos} can be taken out of the sum, which becomes just the sum of Bell polynomials known as the complete Bell polynomial $B_n(a_1,\cdots,a_n)=\sum_{k=1}^n B_{n,k}\left(a_1,\cdots,a^{(n-k+1)} \right)$. Therefore, for a generating function of the form $\Pi_n=e^{g(s)}$, the corresponding probability for the random process is:
\begin{equation}
P(n)=\frac{1}{n!} \frac{d^n}{\partial s^n}\Pi_n|_{s=0} = \frac{e^{g(0)}}{n!} B_n(a_1,\cdots,a_n)
\label{eq:pn}
\end{equation}
with $a_n = g^{(n)}(0)$, and $B_0(\{\})=1$. This reduces the problem of obtaining the photon counting distribution for a combination of random processes to express the generating function as a single exponential $e^{g(s)}$ and computing the $n$-th derivative of the exponent $g^{(n)}(0)$. 

\begin{figure}
\begin{center}
\includegraphics[width=0.9\textwidth]{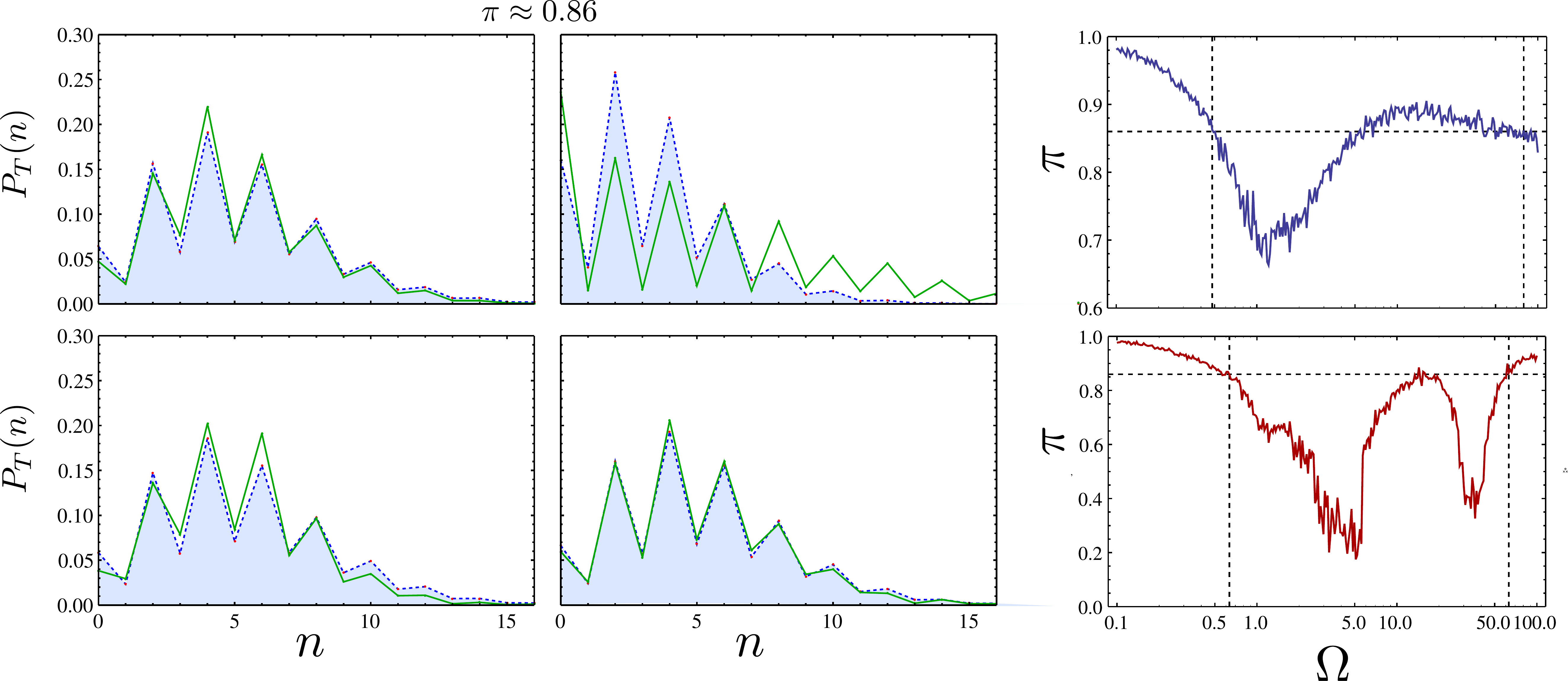}
\end{center}
\caption{Photon counting distributions with the same purity according to some ansatz (in this case, coherent single and two photon processes) are sometimes badly fit. Upper row: $\Delta_\mathrm{I}^\mathrm{2P}$. Lower row: $\Delta_{\mathrm{II}}^{\mathrm{2P}}$. The case of higher $\Omega$ for $\Delta_\mathrm{I}^\mathrm{2P}$ shows the ansatz is not correct in this case. }
\label{fig:fits}
\end{figure}

\section{Thermal and coherent distributions}
The approach to quantify the fraction of two-photon emission requires the selection of a proper ansatz for the photon counting distributions of single and two-photon processes, see Fig.~\ref{fig:fits}. For the case under discussion, we have obtained good results by considering a coherent one-photon distribution, a coherent two-photon distribution and a thermal two-photon distribution. The thermal component is essential to describe the photon counting distributions along the leapfrog line $\Delta_\mathrm{I}^\mathrm{2P}$, giving incorrect results otherwise (see Fig.~\ref{fig:fits}). The generating functions for a coherent one-photon process ($\lambda_1$), a coherent two-photon process ($\lambda_2$) and a thermal two-photon process ($\theta_2$) are given by:
\begin{eqnarray}
\Pi_{\lambda_1} &=& e^{-\lambda_1(1-s)} \\
\Pi_{\lambda_2} &=& e^{-\lambda_1(1-s^2)} \\
\Pi_{\theta_2} &=& \frac{1-\theta_2}{1-s^2\theta_2} = e^{\log{\left(\frac{1-\theta_2}{1-s^2 \theta_2}\right) }}
\end{eqnarray}
This give the total generating function $\Pi_{\lambda_1,\lambda_2,\theta_2} = \Pi_{\lambda_1}  \Pi_{\lambda_2}   \Pi_{\theta_2} = e^{g(s)} $, with
\begin{equation}
g(s) = -\lambda_1(1-s) - \lambda_2(1-s^2) + \log\left(\frac{1-\theta_2}{1-s^2 \theta_2} \right)
\end{equation}
whose $n$-th derivatives are:
\begin{eqnarray}
g^{(n)}(0) = \delta_{n,1}\,  \lambda_1 + \delta_{n,2}\, 2\lambda_2 + \begin{cases}
\theta_2^{n/2}2[(n-1)!] \quad \text{$n$ even}\\
0 \quad \text{$n$ odd}
 \end{cases}
 \label{eq:ansatz}
\end{eqnarray}
This is all the information one needs to construct the photon counting probability given by Eq.~\eqref{eq:pn}.

\section{Thermal and coherent components of the purity}
By fitting Monte Carlo photon counting curves such as the ones shown in Fig.~\ref{fig:fits} one can obtain values for the parameters $\lambda_1$, $\lambda_2$ and $\theta_2$ for a given time window $T$. The mean values associated with each of the three processes are $\bar{n}_{\lambda_1} = \lambda_1$, $\bar{n}_{\lambda_2} = \lambda_2$ and $\bar{n}_{\theta_2} = \theta_2/(1-\theta_2)$. The purity can then be defined as the fraction of the total mean value which is given by two-photon processes, that is:
\begin{equation}
\pi = \frac{\bar{n}_{\lambda_2} + \bar{n}_{\theta_2}}{\bar{n}_{\lambda_1} + \bar{n}_{\lambda_2} + \bar{n}_{\theta_2}} = \frac{\theta_2/(1-\theta_2)+\lambda_2}{\lambda_1+\lambda_2+\theta_2/(1-\theta_2)}.
\label{eq:purity-new}
\end{equation}

This purity can be divided in a thermal plus a coherent part, $\pi = \pi_\theta + \pi_\lambda$, given by:
\begin{eqnarray}
\pi_\theta &=&\frac{\theta_2/(1-\theta_2)}{\lambda_1+\lambda_2+\theta_2/(1-\theta_2)}\\
\pi_\theta &=&\frac{\lambda_2}{\lambda_1+\lambda_2+\theta_2/(1-\theta_2)}
\end{eqnarray}

\begin{figure}
\begin{center}
\includegraphics[width=0.6\textwidth]{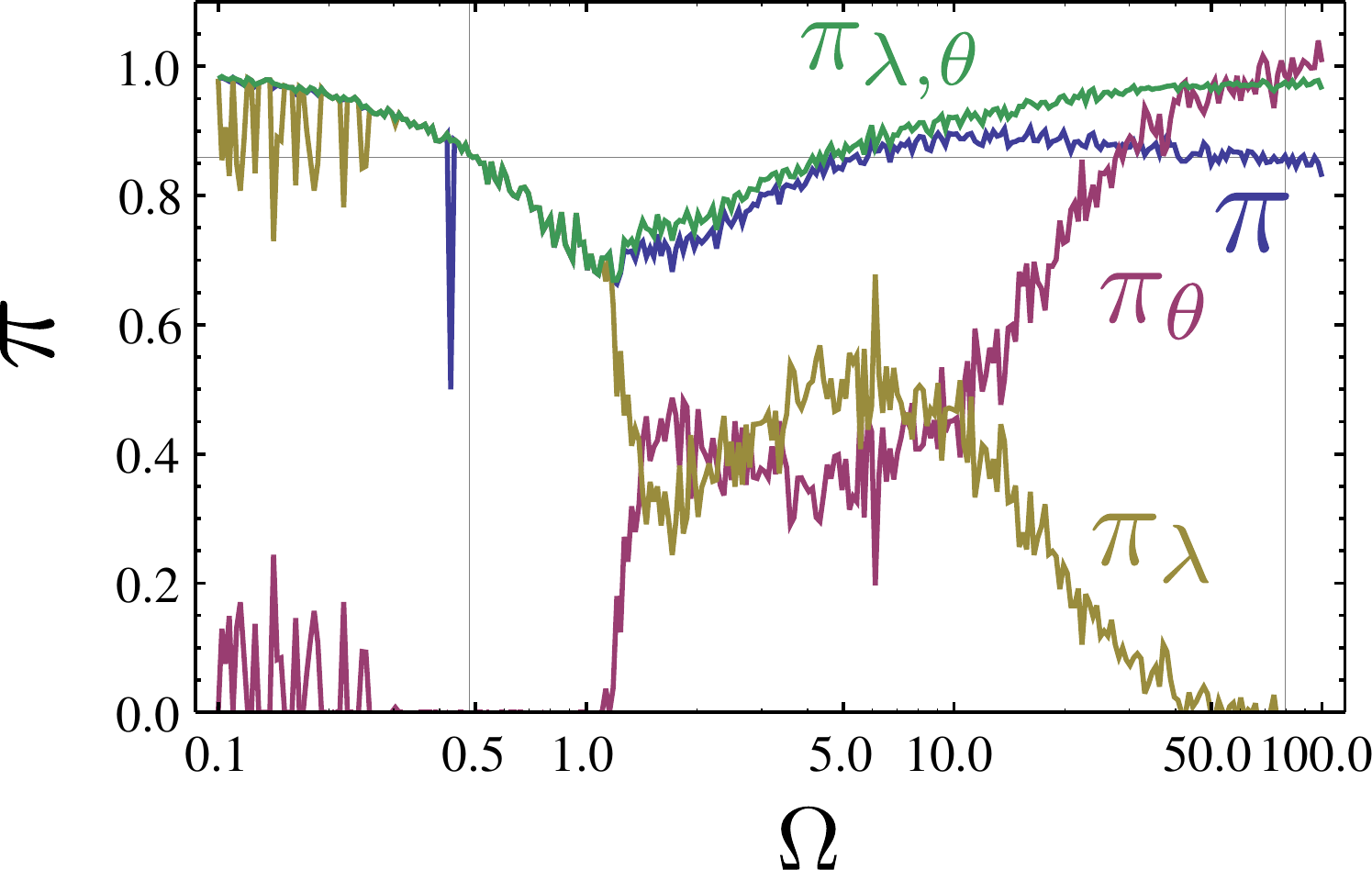}
\end{center}
\caption{Purity for the central leapfrog as calculated by the standard method ($\pi$) and by the cothermal ansantz ($\pi_{\lambda,\theta}$), also with the separated thermal and coherent contributions. We see how the thermal contribution becomes important in this case and gives a higher purity than the one obtained with the standard, coherent ansatz.}
\label{fig:purity-cothermal}
\end{figure}

\begin{figure}
\begin{center}
\includegraphics[width=0.6\textwidth]{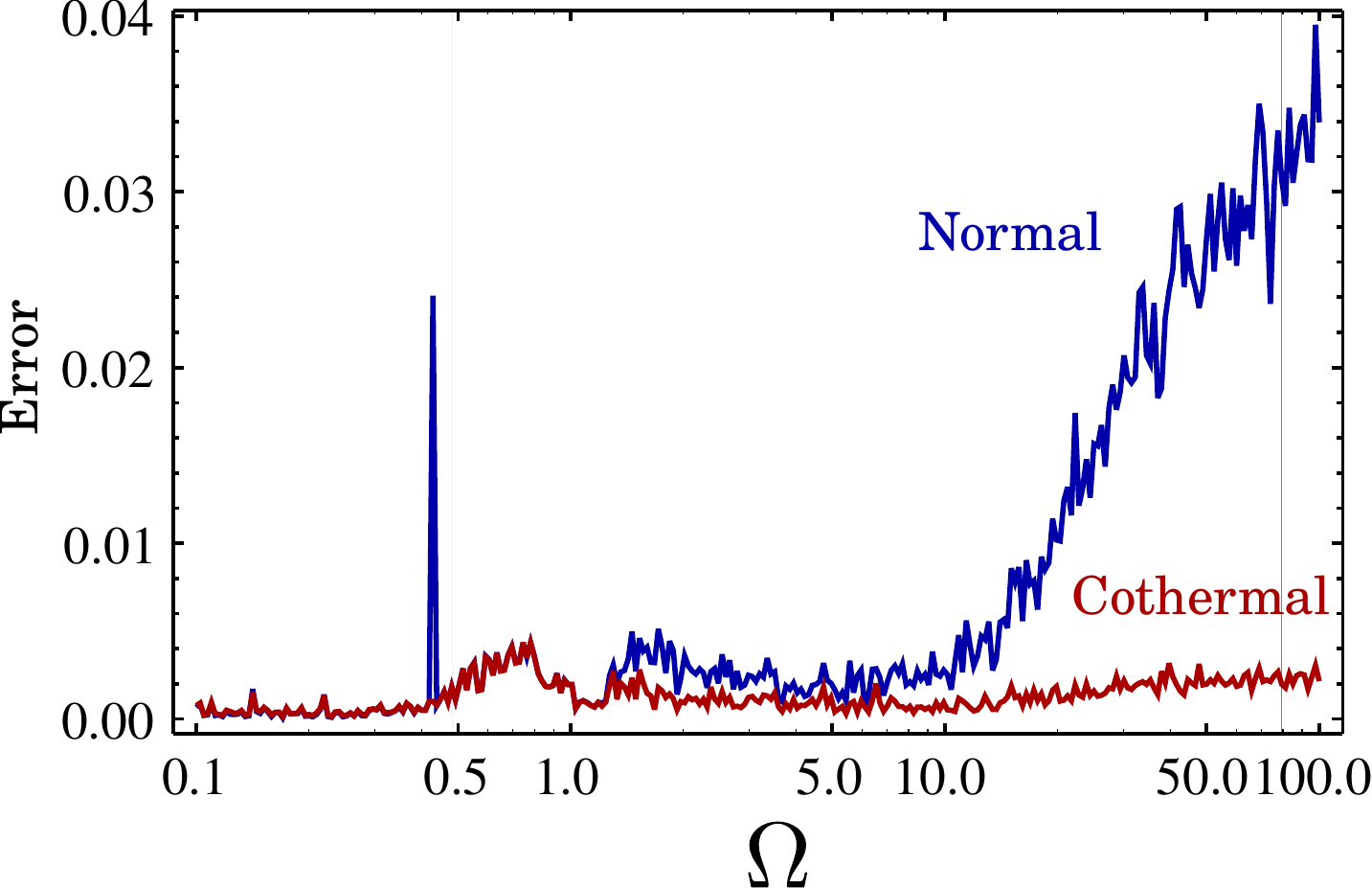}
\end{center}
\caption{Error of the fitting procedure for the central leapfrog when using the standard coherent ansatz (blue) and the cothermal ansatz (red). This shows that the second method gives a much better fitting. }
\label{fig:error-cothermal}
\end{figure}

The result of using the cothermal ansantz \eqref{eq:ansatz} for the photon counting distribution and computing the purity as given by \eqref{eq:purity-new} for the central leapfrog $\Delta_\mathrm{I}^\mathrm{2P}$ in the biexciton configuration is summarized in Fig.~\ref{fig:purity-cothermal}. This provides the expected result of the purity being corresponding to that of a thermal process of bundle emission, which matches the bunched values of $g^{(2)}_2(0)$ in this line. Figure~\ref{fig:error-cothermal} shows the associated errors in the fittings, demonstrating that the cothermal fit provides much better results. 

\end{document}